\documentclass[conference]{IEEEtran}
\usepackage{caption}
\IEEEoverridecommandlockouts
\usepackage{cite}
\usepackage{amsmath,amssymb,amsfonts}
\usepackage{verbatim}
\usepackage{algorithm}      
\usepackage{algorithmic}
\usepackage{graphicx}
\usepackage{textcomp}
\usepackage{xcolor}
\usepackage{float}
\usepackage{bm}
\usepackage{soul}
\usepackage{subcaption}
\usepackage{stfloats}
\usepackage{siunitx}
\usepackage{tikz}
\usetikzlibrary{positioning,arrows.meta,fit,calc}
\usepackage{graphicx}
\usepackage{multirow}

\def\BibTeX{{\rm B\kern-.05em{\sc i\kern-.025em b}\kern-.08em
    T\kern-.1667em\lower.7ex\hbox{E}\kern-.125emX}}

\newcommand{\eqcontrib}{\textsuperscript{\(\circ\)}}

\begin{document}

\title{Uncertainty-Weighted Multi-Task CNN for Joint DoA and Rain-Rate Estimation Under Rain-Induced Array Distortions}

\author{
\IEEEauthorblockN{
    Chenyang Yan\IEEEauthorrefmark{1}\eqcontrib,
    Ruonan Yang\IEEEauthorrefmark{2}\eqcontrib,
    Shunqiao Sun\IEEEauthorrefmark{2},
   and Mats Bengtsson\IEEEauthorrefmark{1}
}
\IEEEauthorblockA{
\eqcontrib\,These authors contributed equally to this work.\\
\IEEEauthorrefmark{1}KTH Royal Institute of Technology, Stockholm, Sweden\\
\IEEEauthorrefmark{2}The University of Alabama, Tuscaloosa, AL USA
}
\thanks{\eqcontrib\,Chenyang Yan and Ruonan Yang contributed equally to this work (co-first authors).}
\thanks{The work of R. Yang and S. Sun was supported in part by U.S. National Science Foundation (NSF) under Grants CCF-2153386 and ECCS-2033433. The computations were partly enabled by resources provided by the National Academic Infrastructure for Supercomputing in Sweden (NAISS), partially funded by the Swedish Research Council through grant agreement no. 2022-06725.}

}

\maketitle

\begin{abstract}
We investigate joint direction-of-arrival (DoA) and rain-rate estimation for a uniform linear array operating under rain-induced multiplicative distortions. Building on a wavefront fluctuation model whose spatial correlation is governed by the rain-rate, we derive an angle-dependent covariance formulation and use it to synthesize training data. DoA estimation is cast as a multi-label classification problem on a discretized angular grid, while rain-rate estimation is formulated as a multi-class classification task. We then propose a multi-task deep CNN with a shared feature extractor and two task-specific heads, trained using an uncertainty-weighted objective to automatically balance the two losses. Numerical results in a two-source scenario show that the proposed network achieves lower DoA RMSE than classical baselines and provides accurate rain-rate classification at moderate-to-high SNRs.
\end{abstract}


\section{Introduction}
\label{sec:1}

Direction-of-arrival (DoA) estimation is a key enabler for radar-based autonomous driving and obstacle detection~\cite{sun2020mimo,kulkarni2025comprehensive,jekaterynczuk2023survey}. Under heavy rain, propagation effects such as attenuation and backscattering distort array measurements and can substantially degrade DoA performance, especially in multi-source localization~\cite{zang2019impact,Ballistic}.

In our previous work, we developed a physics-based rain-impaired array model and a covariance-matching calibration method~\cite{yan2025robustcovariance} based on~\cite{yektakhah2024model}. However, the distortion should be angle-dependent and the analysis was restricted to the single-source case. Moreover, since the distortion statistics are governed by the rain-rate, jointly estimating the rain-rate alongside DoA is both feasible and practically useful as an auxiliary environmental indicator.

Motivated by recent learning-based DoA estimators that show improved robustness under low SNR, model mismatch, and complex environments~\cite{liu2020super,papageorgiou2021deep,merkofer2023music, Model_based_learning_2023, Yang2025Advancing,hu2025model}, we propose a multi-task deep CNN that learns DoA and rain-rate information directly from covariance measurements under rain-induced distortions.

The contributions of this paper are summarized as follows:
\begin{itemize}
    \item We propose a CNN-based framework for joint DoA and rain-rate estimation under a rain-impaired array measurement model, and validate it in a two-source scenario.
    \item We develop an uncertainty-weighted multi-task objective that automatically balances the DoA and rain-rate classification losses during training.
    \item Experiments under rain-induced distortions demonstrate the potential advantages of deep learning--based joint estimation over classical baselines in challenging conditions.
\end{itemize}

\section{Rain-Induced Distortion and Signal Model}\label{sec:2}
This section presents the rain-induced distortion model and its resulting array signal formulation. We first describe the wavefront-level fluctuation statistics of the complex electric field and introduce an empirical correlation model parameterized by the rain-rate. We then map the wavefront distortions to a multiplicative array observation model and extend it to the multi-source ULA case with angle-dependent distortions. Finally, we lift the model to the covariance domain, where the rain-rate information is embedded in the distortion covariance $\mathbf{R}_b(\theta)$ through a correlation coefficient $\alpha$, providing the basis for subsequent joint DoA and rain-rate estimation.

\subsection{Rain-Induced Distortion Model}\label{sec:rain_model}

Following the rain-induced wavefront fluctuation model in ~\cite{yektakhah2024model}, let $P$ denote a point on the received wavefront. For a given rain realization, let $E(P)\in\mathbb{C}$ denote the complex electric-field envelope at $P$. The mean (coherent) field is defined as $E_\text{mean}(P)\triangleq \mathbb{E}\{E(P)\}$, where $\mathbb{E}\{\cdot\}$ denotes the ensemble average over independent rain realizations. The normalized electric-field fluctuation is defined as
\begin{equation}
E_n(P) \triangleq \frac{E(P)-E_\text{mean}(P)}{E_\text{mean}(P)}.
\label{eq:normalized_field}
\end{equation}

Consider two points $P_1$ and $P_2$ on the same wavefront with separation $d$, and define $E_{n,1}\triangleq E_n(P_1)$ and $E_{n,2}\triangleq E_n(P_2)$. We model $E_{n,1}$ and $E_{n,2}$ as zero-mean, jointly circularly symmetric complex Gaussian random variables with second-order statistics:
\begin{align}
\mathbb{E}\{|E_{n,1}|^2\} &= \mathbb{E}\{|E_{n,2}|^2\} = 2\lambda_{11}, \\
\mathbb{E}\{E_{n,1} E_{n,2}^*\} &= 2\alpha\lambda_{11}, \quad 
\Im\!\left\{\mathbb{E}\{E_{n,1} E_{n,2}^*\}\right\}=0,
\end{align}
where $\lambda_{11}$ controls the fluctuation power and $\alpha$ characterizes the spatial correlation on the wavefront. The parameter $\alpha$ depends on the rain-rate, propagation distance, separation $d$, and radar carrier frequency, and can be computed using the empirical model in Eq.~(13) of~\cite{yektakhah2024model}:
\begin{equation}
\alpha = \exp\!\left(
-a_1 \left(\frac{R}{a_2 R + 1}\right)
\left(\frac{\tfrac{d}{\lambda_0}}{a_3 \tfrac{d}{\lambda_0} + 1}\right)
\right),
\label{eq:empirical_alpha}
\end{equation}
where $R$ denotes the propagation range, $\lambda_0$ is the wavelength, and $a_1$, $a_2$, and $a_3$ are parameters determined by the rain-rate and carrier frequency (see Table~II in~\cite{yektakhah2024model}). According to~\cite{yektakhah2024model}, \eqref{eq:empirical_alpha} is valid for $R \leq 500~\mathrm{m}$ and $0.1 \leq d/\lambda_0 \leq 8$. The probability density functions (pdfs) of the phase difference and magnitude ratio under varying parameters (e.g., separation $d$, range, and rain-rate) are illustrated in Figs.~1 and 2 of~\cite{yan2025robustcovariance}.

\subsection{Array Observation Model}\label{sec:signal_model}

We now connect the wavefront model to array observations. Consider a uniform linear array (ULA) with $M$ sensors. Let $P_m$ denote the location of the $m$th sensor, and identify the received signal $y_m(t)$ with the complex electric field $E(P_m,t)$. Substituting \eqref{eq:normalized_field} yields
\begin{equation}
y_m(t) = \bigl(1 + E_n(P_m,t)\bigr)\,E_\text{mean}(P_m),
\label{eq:ym_rain}
\end{equation}
where $E_\text{mean}(P_m)$ denotes the mean field component, i.e., the distortion-free received field at $P_m$. Accordingly, we define the multiplicative distortion at sensor $m$ as
\begin{equation}
b_m(t) \triangleq 1 + E_n(P_m,t).
\label{eq:b_m}
\end{equation}
Note that this definition differs from our previous implementation in~\cite{yan2025robustcovariance}, where the distortion term was taken as $b_m(t)=E_n(P_m,t)$. This difference corresponds to a reparameterization that absorbs the constant offset into the distortion coefficient; therefore, the subsequent modeling and the proposed algorithm remain applicable under~\eqref{eq:b_m}.

In~\cite{yan2025robustcovariance}, we introduced a rain-impaired array reception model in which the complex electric-field fluctuations occur across the incident plane wave as it reaches the ULA aperture (see Fig.~3 in~\cite{yan2025robustcovariance}). For the single-source case, the corresponding covariance model was derived in Eq.~(9) of~\cite{yan2025robustcovariance}, where the rain-induced distortion appears as a Hadamard product between the distortion-free covariance matrix and a real Toeplitz distortion matrix $\mathbf{R}_b$, defined in Eq.~(10) of~\cite{yan2025robustcovariance}.

Starting from Eq.~(5) in~\cite{yan2025robustcovariance}, we extend the model to $N$ narrowband sources, where the $n$th source impinges from direction $\theta_n$. The measurement model of a rain-distorted ULA is then given by
\begin{equation}
\mathbf{y}(t) = \sum_{n=1}^{N} \Bigl[\mathbf{a}(\theta_n) \odot \mathbf{b}(\theta_n,t)\Bigr] s_n(t) + \mathbf{n}(t),
\label{eq:multi_measurement}
\end{equation}
where $\mathbf{y}(t)\in\mathbb{C}^{M}$ denotes the received array snapshot, $\mathbf{a}(\theta_n)\in\mathbb{C}^{M}$ is the distortion-free steering vector, $s_n(t)\in\mathbb{C}$ is the $n$th source signal, and $\mathbf{n}(t)\in\mathbb{C}^{M}$ denotes additive noise. The vector $\mathbf{b}(\theta_n,t)\in\mathbb{C}^{M}$ represents the multiplicative rain-induced distortion affecting the $n$th wavefront across the array and is defined as
\begin{equation}
\mathbf{b}(\theta_n,t) \triangleq 
\begin{bmatrix}
b_{n,1}(t),\, b_{n,2}(t),\, \dots,\, b_{n,M}(t)
\end{bmatrix}^{T},
\label{eq:b_vec}
\end{equation}
where $b_{n,m}(t)$ denotes the distortion coefficient at sensor $m$ associated with the wavefront arriving from direction $\theta_n$.

The distortion vector $\mathbf{b}(\theta_n,t)$ is angle-dependent because the effective separation on the reference wavefront plane (see Fig.~3 in~\cite{yan2025robustcovariance}) varies with the incidence angle. In particular, for an inter-element spacing $d_0$ in wavelength along the array axis, the corresponding inter-element spacing in wavelength on the wavefront is
\begin{equation}
d(\theta_n) = d_0 \cos(\theta_n).
\label{eq:d_theta}
\end{equation}
Hence, the correlation parameter $\alpha$ (and thus the statistics of $\mathbf{b}(\theta_n,t)$) inherits an explicit dependence on $\theta_n$ through $d(\theta_n)$.

\subsection{Covariance-Domain Formulation}\label{sec:cov_model}

Alternatively, \eqref{eq:multi_measurement} can be lifted to the covariance domain. Define the source vector $\mathbf{s}(t)\triangleq[s_1(t),\dots,s_N(t)]^{T}\in\mathbb{C}^{N}$ and the DoA collection $\boldsymbol{\theta}\triangleq[\theta_1,\dots,\theta_N]^{T}$. We assume that $\mathbf{s}(t)$ and $\mathbf{n}(t)$ are independent, zero-mean, circularly symmetric complex processes. Moreover, we assume mutually uncorrelated sources with equal power, such that
\begin{equation}
\mathbf{R}_s \triangleq \mathbb{E}\{\mathbf{s}(t)\mathbf{s}(t)^{H}\} = \sigma_s^2 \mathbf{I}_N, \
\mathbf{R}_n \triangleq \mathbb{E}\{\mathbf{n}(t)\mathbf{n}(t)^{H}\},
\end{equation}
where $\sigma_s^2$ denotes the per-source signal power and $\mathbf{I}_N$ is the $N\times N$ identity matrix. Under these assumptions, the covariance matrix of $\mathbf{y}(t)$ is given by
\begin{equation}
\begin{split}
\mathbf{R}_y &\triangleq \mathbb{E}\{\mathbf{y}(t)\mathbf{y}(t)^{H}\} \\
&= \sum_{n=1}^{N} \Bigl[\mathbf{R}_x(\theta_n)\odot \mathbf{R}_b(\theta_n)\Bigr] + \mathbf{R}_n,
\end{split}
\label{eq:covariance_measurement}
\end{equation}
where $\mathbf{R}_x(\theta_n)$ denotes the distortion-free array covariance contribution of the $n$th source,
\begin{equation}
\mathbf{R}_x(\theta_n) = \sigma_s^2\,\mathbf{a}(\theta_n)\mathbf{a}(\theta_n)^{H}.
\label{eq:Rx_def}
\end{equation}
The matrix $\mathbf{R}_b(\theta_n)\in\mathbb{C}^{M\times M}$ is the rain-induced distortion covariance associated with direction $\theta_n$, defined elementwise as
\begin{equation}
\bigl[\mathbf{R}_b(\theta_n)\bigr]_{m\ell} \triangleq \mathbb{E}\bigl\{ b_{n,m}(t)\, b_{n,\ell}^*(t) \bigr\}, \ 1\le m,\ell\le M.
\label{eq:Rb_def}
\end{equation}
Using $b_{n,m}(t)=1+E_n(P_m,t)$ and the second-order statistics of $E_n(\cdot)$, we obtain
\begin{equation}
\bigl[\mathbf{R}_b(\theta_n)\bigr]_{m\ell} =
\begin{cases}
1 + 2\lambda_{11}, & m=\ell,\\[2pt]
1 + 2\lambda_{11}\,\alpha_{m\ell}(\theta_n), & m\neq \ell,
\end{cases}
\label{eq:Rb_piecewise}
\end{equation}
where $\alpha_{m\ell}(\theta_n)$ denotes the correlation parameter corresponding to the sensor-pair separation. In particular, $\alpha_{m\ell}(\theta_n)$ is evaluated by~\eqref{eq:empirical_alpha} with
\begin{equation}
d = |m-\ell|\,d(\theta_n), \qquad d(\theta_n)=d_0\cos(\theta_n).
\label{eq:d_mell}
\end{equation}
Hence, $\mathbf{R}_b(\theta_n)$ inherits an explicit dependence on $\theta_n$ through the angle-dependent separation $d(\theta_n)$. Moreover, since $\alpha_{m\ell}(\theta_n)$ is computed via~\eqref{eq:empirical_alpha}, the rain-rate information is fully captured by $\mathbf{R}_b(\theta_n)$ (through $a_1$, $a_2$, and $a_3$) and propagates to the observed covariance $\mathbf{R}_y$ in~\eqref{eq:covariance_measurement}.

\section{Proposed Method}\label{sec:method}
In this section, we formulate DoA estimation as a multi-label classification task and rain-rate estimation as a multi-class classification task. We then develop a multi-task CNN architecture that jointly learns these tasks from the sample covariance matrix $\mathbf{R}_y$, thereby exploiting both DoA- and rain-rate-dependent information embedded in the covariance measurements.

\subsection{Input Representation and Preprocessing}\label{sec:input}
Following~\cite{papageorgiou2021deep}, we cast DoA prediction as a multi-label classification task over a discretized angular grid. Specifically, the search interval $[-\theta_{\max},\theta_{\max}]$ is uniformly quantized with resolution $\rho$ to form the grid $\mathcal{G}$. The DoA label is encoded as a binary vector over $\mathcal{G}$, where the entries corresponding to the true source angles are set to one and all other entries are set to zero.

For rain-rate classification, we consider six discrete classes. Following Table~II in~\cite{yektakhah2024model}, we use five rain-rate levels $\{2,5,10,25,50\}$~mm/h and add an additional no-rain class (0~mm/h), for which no rain-induced distortion is applied (i.e., the distortion coefficients are set to unity). For each rain-rate class, the corresponding distortion covariance is constructed from~\eqref{eq:Rb_piecewise}. We then synthesize spatially correlated complex Gaussian distortion samples by transforming i.i.d.\ Gaussian draws with a Cholesky factor of the target covariance (a square-root covariance transformation)~\cite{hammersley2013monte}. These samples are used to generate rain-distorted covariance measurements for training.

For each training example, we compute the sample covariance matrix $\hat{\mathbf{R}}_y$ from the received snapshots and construct a real-valued three-channel input tensor $\mathbf{X}\in\mathbb{R}^{M\times M\times 3}$, whose channels are $\Re\{\hat{\mathbf{R}}_y\}$, $\Im\{\hat{\mathbf{R}}_y\}$, and $\angle\{\hat{\mathbf{R}}_y\}$, respectively~\cite{papageorgiou2021deep}. Here, $\Re\{\cdot\}$ and $\Im\{\cdot\}$ denote the real and imaginary parts of a complex matrix, and $\angle\{\cdot\}$ denotes the phase (argument) of a complex-valued entry.

\subsection{Network Architecture}\label{sec:arch}
The nonlinear mapping $f(\cdot)$ is parameterized by a deep convolutional neural network (CNN) comprising a shared feature extractor and task-specific prediction heads. Let $\mathbf{X}\in\mathbb{R}^{M\times M\times 3}$ denote the input tensor constructed from the sample covariance matrix, where the three channels contain the real part, imaginary part, and phase, respectively. The network first applies a cascade of four convolutional blocks:
\begin{equation}
f(\mathbf{X}) = f_{4}\!\bigl(f_{3}\!\bigl(f_{2}\!\bigl(f_{1}(\mathbf{X})\bigr)\bigr)\bigr).
\end{equation}
These four convolutional blocks collectively constitute a shared CNN encoder. Each block $f_i(\cdot)$, $i\in\{1,2,3,4\}$, consists of a $2$-D convolution with $n_C=256$ filters, followed by batch normalization and a rectified linear unit (ReLU). All convolutions use a $3\times 3$ kernel with stride $1$ and padding $1$. A $2\times 2$ max-pooling layer with stride $2$ is inserted after the second block $f_{2}(\cdot)$. The output feature maps after $f_4(\cdot)$ are then flattened into a feature vector.

\begin{figure}
     \centering
     \includegraphics[width=\linewidth]{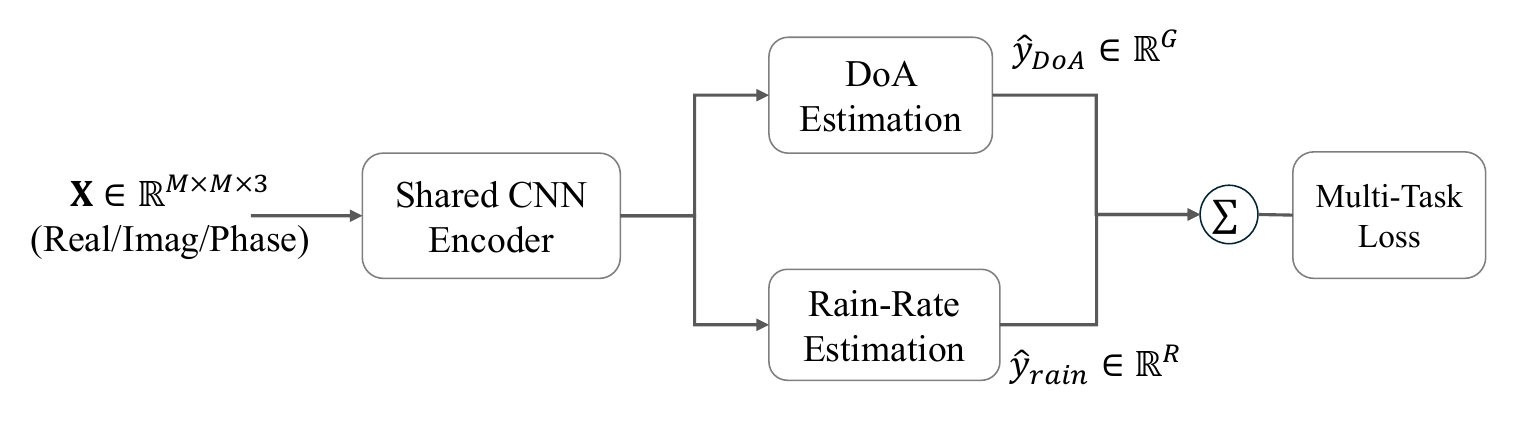}
     \caption{ Proposed multi-task network for joint DoA and rain-rate estimation. The input tensor $\mathbf{X}\in\mathbb{R}^{M\times M\times 3}$ stacks the real, imaginary, and phase channels of the sample covariance matrix and is processed by a shared CNN encoder. Two task-specific heads output DoA logits $\hat{\mathbf{y}}_{\mathrm{DoA}}\in\mathbb{R}^{G}$ and rain-rate logits $\hat{\mathbf{y}}_{\mathrm{rain}}\in\mathbb{R}^{R}$, which are jointly optimized via a multi-task loss.
 }
     \label{fig:cnn_arch}
 \end{figure}

The resulting feature vector is fed to a fully connected layer with $256$ hidden units and ReLU activation, followed by a linear output layer that produces the DoA logits $\hat{\mathbf{y}}_{\mathrm{DoA}}\in\mathbb{R}^{G}$, where $G$ is the number of angular grid points. In parallel, a task-specific rain-rate head with the same two-layer fully connected structure outputs the rain-rate logits $\hat{\mathbf{y}}_{\mathrm{rain}}\in\mathbb{R}^{R}$, where $R$ denotes the number of rain-rate classes. During training, the DoA logits are converted to independent Bernoulli probabilities via a sigmoid and optimized using a multi-label binary cross-entropy (BCE) loss, whereas the rain-rate logits are converted to a categorical distribution via a softmax and optimized using cross-entropy (CE). In practice, these nonlinearities are applied implicitly within the respective loss functions.
Finally, we associate each task with a learnable log-variance parameter, which is used for uncertainty-based reweighting of the multi-task objective. The structure of the network can be seen in Fig.~\ref{fig:cnn_arch}

\subsection{Multi-Task Learning with Uncertainty Weighting}\label{sec:mtl}
The proposed network takes the covariance-based input $\mathbf{X}$ and outputs two sets of logits:
$\hat{\mathbf{y}}_{\mathrm{DoA}}\in\mathbb{R}^{G}$ for multi-label DoA prediction over the grid $\mathcal{G}$ and
$\hat{\mathbf{y}}_{\mathrm{rain}}\in\mathbb{R}^{R}$ for multi-class rain-rate prediction.
To automatically balance the two tasks during training, we adopt uncertainty weighting~\cite{kendall2018multi}.

Let $\mathbf{y}_{\mathrm{DoA}}\in\{0,1\}^{G}$ denote the multi-label DoA target and
$y_{\mathrm{rain}}\in\{1,\ldots,R\}$ denote the rain-rate class label.
We associate each task with a learnable scale parameter,
$\sigma_{\mathrm{DoA}}>0$ and $\sigma_{\mathrm{rain}}>0$. We model the $G$ binary labels as conditionally independent Bernoulli variables with scaled logits.
Specifically, defining the element-wise sigmoid as $\mathrm{g}(\cdot)$, we write
\begin{equation}
\begin{aligned}
&p(\mathbf{y}_{\mathrm{DoA}}\mid \mathbf{X},\mathbf{W},\sigma_{\mathrm{DoA}})
=
\prod_{g=1}^{G}
\Big[
\mathrm{g}\!\Big(\hat y_{\mathrm{DoA},g}/\sigma_{\mathrm{DoA}}^{2}\Big)^{y_{\mathrm{DoA},g}}
\\ &\times 
\Big(1-g\!\big(\hat y_{\mathrm{DoA},g}/\sigma_{\mathrm{DoA}}^{2}\big)\Big)^{1-y_{\mathrm{DoA},g}}
\Big].
\end{aligned}
\label{eq:doa_bern_sigmoid}
\end{equation}

The corresponding negative log-likelihood reduces to a binary cross-entropy loss in logit form, where the DoA logits are scaled by $1/\sigma_{\mathrm{DoA}}^{2}$. 
In other words, we compute the multi-label BCE loss using the scaled logits $\hat{\mathbf{y}}_{\mathrm{DoA}}/\sigma_{\mathrm{DoA}}^{2}$ and the target vector $\mathbf{y}_{\mathrm{DoA}}$. The unscaled DoA loss is obtained by evaluating the same BCE on the original logits $\hat{\mathbf{y}}_{\mathrm{DoA}}$. Throughout this section, the scaled loss is expressed as $\mathcal{L}_{\mathrm{task}}(\mathbf{W},\sigma_{\mathrm{task}})$ and the unscaled loss is expressed as $\mathcal{L}_{\mathrm{task}}(\mathbf{W})$

Following the approximation used for classification in~\cite{kendall2018multi}, the effect of temperature scaling
can be expressed as a weighted unscaled loss plus a scale-dependent regularizer. Summing over the $G$ independent
Bernoulli factors yields
\begin{equation}
\mathcal{L}_{\mathrm{DoA}}(\mathbf{W},\sigma_{\mathrm{DoA}})
\approx
\frac{1}{\sigma_{\mathrm{DoA}}^{2}}\,\mathcal{L}_{\mathrm{DoA}}(\mathbf{W})
+ G\log\sigma_{\mathrm{DoA}} .
\label{eq:doa_uncertainty_compact}
\end{equation}

For rain-rate prediction we use the standard softmax cross-entropy loss
$\mathcal{L}_{\mathrm{rain}}(\mathbf{W})=\ell_{\mathrm{CE}}(\hat{\mathbf{y}}_{\mathrm{rain}},y_{\mathrm{rain}})$.
Its uncertainty-weighted form is adopted directly from~\cite{kendall2018multi}:
\begin{equation}
\mathcal{L}_{\mathrm{rain}}(\mathbf{W},\sigma_{\mathrm{rain}})
\approx
\frac{1}{\sigma_{\mathrm{rain}}^{2}}\,\mathcal{L}_{\mathrm{rain}}(\mathbf{W})
+\log\sigma_{\mathrm{rain}} .
\label{eq:rain_uncertainty_compact}
\end{equation}
For numerical stability, we optimize the log-variances
$s_{\mathrm{DoA}}=\log\sigma_{\mathrm{DoA}}^{2}$ and
$s_{\mathrm{rain}}=\log\sigma_{\mathrm{rain}}^{2}$.
Combining~\eqref{eq:doa_uncertainty_compact} and~\eqref{eq:rain_uncertainty_compact} gives
\begin{align}
\mathcal{L}(\mathbf{W},s_{\mathrm{DoA}},s_{\mathrm{rain}})
=
& e^{-s_{\mathrm{DoA}}}\mathcal{L}_{\mathrm{DoA}}(\mathbf{W})  +e^{-s_{\mathrm{rain}}}\mathcal{L}_{\mathrm{rain}}(\mathbf{W}) \nonumber\\
&+\frac{G}{2}s_{\mathrm{DoA}}+\frac{1}{2}s_{\mathrm{rain}} . \label{eq:mtl_uncertainty_final_compact}
\end{align}

\section{Evaluation and Results}
\label{sec:4}

For training, we discretize the DoA space using a $1^{\circ}$ grid over a field of view (FoV) of $60^{\circ}$ centered at $0^{\circ}$, i.e., $\mathcal{G}=\{-30^{\circ},-29^{\circ},\ldots,30^{\circ}\}$, which yields a grid size of $G=61$. We consider a fixed two-source scenario ($N=2$) and fix the propagation range to $R=200$~m. We then generate covariance samples uniformly across the six rain-rate classes described above and five SNR levels $\{0,5,10,15,20\}$~dB. In total, $100{,}000$ samples are generated for training. The dataset is randomly split into a training set (90\%) and a validation set (10\%).

To evaluate the effect of multi-task loss weighting, we further compare the proposed uncertainty-weighted network with fixed loss weighting using a predefined ratio \(\alpha\), where the overall training loss is given by $\mathcal{L} = \mathcal{L}_{\mathrm{DoA}} + \alpha\,\mathcal{L}_{\mathrm{rain}}$. Results are reported for \(\alpha = 1\) and \(\alpha = 0.05\), enabling a direct comparison between the proposed adaptive weighting and fixed loss-weighting schemes.

\begin{figure}
    \centering
\includegraphics[width=\linewidth]{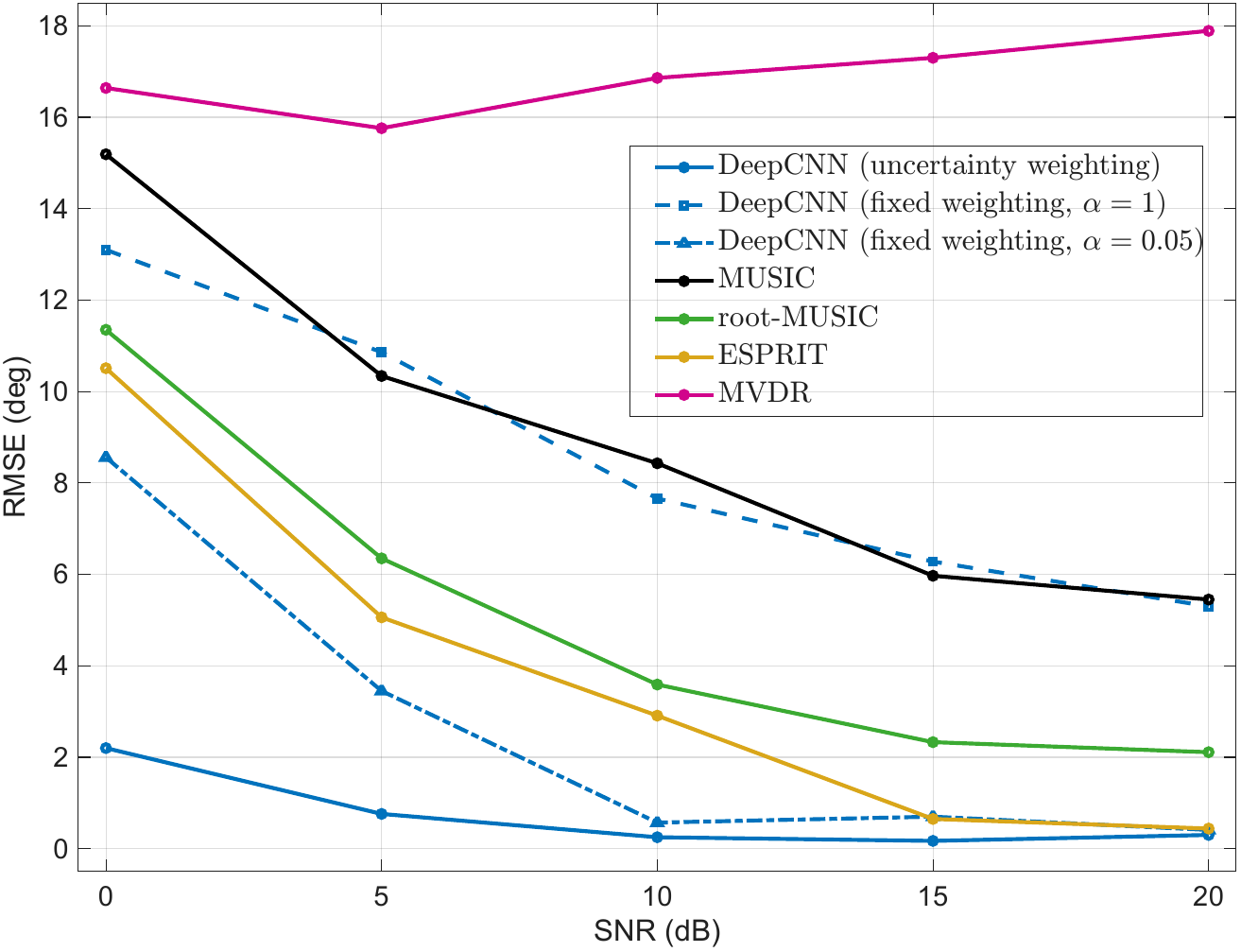}
    \caption{DoA RMSE versus SNR on the test set under rain-induced array distortions. The proposed uncertainty-weighted network is compared with its fixed-weighting variants and classical baselines.}
    \label{rmse}
    \vspace{-3mm}
\end{figure}

Fig.~\ref{rmse} compares the DoA root-mean-square error (RMSE) of the proposed network with classical baselines, including MUSIC, root-MUSIC, ESPRIT, and MVDR,  on the same test set consisting of $10{,}000$ samples. In addition to outperforming the model-based methods, the proposed uncertainty-weighted network consistently improves upon its fixed-weighting variants across all SNRs. The benefit is most evident in the low-SNR regime, where uncertainty weighting yields noticeably lower RMSE, indicating that adaptive task balancing leads to more robust DoA inference under noisy and rain-distorted measurements. Table~\ref{table:rainrate_acc} compares the rain-rate classification accuracy of the proposed uncertainty-weighted network with fixed-weighting variants. The proposed model consistently attains \(\ge 94\%\) rain-rate classification accuracy across all tested SNR levels, indicating reliable rain-rate inference even in low-SNR conditions.
Overall, the proposed uncertainty-weighted network achieves reliable performance on both DoA estimation and rain-rate classification in a unified multi-task framework.

\begin{table}[h]
\centering
\caption{Rain-rate classification accuracy (\%) versus SNR for different loss weighting strategies.}
\label{table:rainrate_acc}
\begin{tabular}{c c c c}
\hline\hline
\multirow{2}{*}{\textbf{SNR (dB)}} 
& \multirow{2}{*}{\textbf{Uncertainty weighting (see~\eqref{eq:mtl_uncertainty_final_compact})}} 
& \multicolumn{2}{c}{\textbf{Fixed weighting}} \\
& & $\boldsymbol{\alpha = 0.05}$ & $\boldsymbol{\alpha = 1}$ \\
\hline
0  & 94.21 & 63.95 & 87.13 \\
5  & 94.55 & 78.09 & 95.19 \\
10 & 96.07 & 92.80 & 99.09 \\
15 & 96.79 & 99.16 & 99.95 \\
20 & 99.41 & 99.85 & 99.95 \\
\hline\hline
\end{tabular}
\end{table}

\vspace{-2mm}

\section{Conclusions}
\label{sec:5}

This paper proposes a CNN-based multi-task framework for jointly estimating the DoA and the rain-rate under rain-induced array distortions. The results provide initial evidence that the proposed formulation is feasible and can outperform classical DoA estimators in the considered setting. In addition, we extend uncertainty-weighted multi-task learning to a multi-label (sigmoid) DoA loss with a multi-class (softmax) rain-rate loss, yielding a principled and learnable mechanism to balance the two objectives.

Future work will validate the approach using real-world radar measurements, extend the current discrete rain-rate classifier to a continuous (regression) formulation, and investigate more challenging scenarios with a larger (and possibly varying) number of simultaneously active sources, including cases where the source count is unknown.

\newpage
\bibliographystyle{IEEEtran}
\bibliography{ref}

\end{document}